\documentstyle[epsfig,12pt]{article}
\input{epsf.sty}
\newcommand{\pp}{{_{I\hspace{-0.2em}P}}}
\begin{document}
\parskip 0.3cm
\begin{titlepage}
\begin{flushright}

\end{flushright}
\vspace{1.cm}

\begin{centering}
{\large \bf Diffractive Photoproduction in the Framework of Fracture Functions.}

 \vspace{.9cm}
{\bf R. Sassot$^{\dagger}$} \\
\vspace{.05in}
{\it Departamento de F\'{\i}sica, Universidad de Buenos Aires \\
Ciudad Universitaria, Pab.1 (1428) Buenos Aires, Argentina} \\
\vspace{1.5cm}
{ \bf Abstract} \\
\bigskip
\end{centering}
{\small Recent data on diffractive photoproduction of dijets are
analyzed within the framework of fracture functions and paying special
attention to the consequences of the use of different rapidity gap
definitions in order to identify diffractive events. Although these
effects are found to be significant, it is shown that once they
are properly taken into account, a very precise agreement between
diffractive DIS and diffractive dijet photoproduction emerges without
any significant hint of hard factorization breaking. \\ \\
13.60.-r 13.87.-a 13.85.Ni}\\

\vspace*{30mm}
\noindent$^{\dagger}$ Fellow of CONICET, Argentina.

\end{titlepage}

\noindent{\large \bf Introduction}\\
 
Recent years have seen a renewed interest in diffractive physics and
specifically, in the possibility of using hard diffractive processes
in order to investigate `at parton level' diverse features of
diffraction \cite{INGDON}.

The success of Regge theory in the phenomenological description of
soft diffraction \cite{GOU83} has naturally encouraged most analyses 
to use not
only the language but also many critical hypothesis of that approach
but in the hard regime and including simultaneously elements of
perturbative QCD\cite{INGDON,BER,KUN}. In this way, issues like the `parton
 content of the
pomeron' or the `factorization properties of the pomeron parton
densities' have become the center of active experimental programs and
intense theoretical debates \cite{ABR}.

This concoction of perturbative QCD and Regge ideas has shown to be
quite successful, from the phenomenological point of view, when applied
individually to an increasingly large set of diffractive deep
inelastic scattering data \cite{H1D,ZD}, at least as a first approximation, 
and also
to diffractive processes within hadron-hadron collisions \cite{TEV}. However,
 it seems to have serious problems when it is meant as a processes independent
description of both sets of data \cite{COLLB}. This fact could indicate a
breakdown either of universality in the ingredients coming form the Regge
picture, due for instance to an unexpected energy dependence in the
pomeron probability density \cite{GOUE}, or in the hard factorization property assumed in
the QCD inspired component of the treatment \cite{COLLB,COLLS,BER}.

Alternatively to the above mentioned approach, many hard diffractive
processes, including diffractive DIS, can be treated rigorously within
perturbative QCD -and without caring about the validity of the
hypothesis implicit in the Regge approach- using the framework of
fracture functions \cite{ph}. This framework was originally thought to deal with
semi-inclusive processes within QCD in a fully consistent way \cite{VEN}, and
more recently it has been shown to be particularly successful in the
phenomenological description of diffractive DIS and leading baryon
production, and also in the analysis of departures from the standard Regge
hypothesis, already found in the available data \cite{ph}.

In the present paper, the framework of fracture functions is applied
to diffractive photoproduction of dijets \cite{Zph,Hph}. This is meant not only as a
test of the general approach and as a complementary verification of
the parameterizations extracted from DIS, but also as a test of the
hard factorization property assumed for fracture functions in the case
 of hadron-hadron collisions. At variance with the case of diffractive
DIS \cite{COLLF}, the factorization property of fracture functions, which is crucial 
in order to compute the sizeable contribution of resolved photons to the
photoproduction processes \cite{Zphv,Zph,Hph},
may be challenged by the occurrence of soft interactions before the hard scattering takes place in diffractive hadron-hadron collisions \cite{BER,COLLS}.

Performing this analysis, it has been found that the use of
non-equivalent rapidity gap definitions in order to identify
diffractive processes in the different experiences conspires strongly
against the above mentioned tests. In the language of fracture
functions, to adopt different gap definitions implies that only
certain components of the whole fracture function are tested,
suppressing or enhancing contributions coming from certain kinematical
regions,
and leading seemingly to a  factorization breakdown. However, here it
is shown that these effects can be properly taken into account
yielding a very precise agreement between diffractive DIS and
diffractive dijet photoproduction, without any significant hint of
hard factorization breaking.

In the following section, the standard procedure used in the
extraction of diffractive parton densities or, alternatively, fracture
functions in diffractive DIS is reviewed giving emphasis to the issue
of the use of different definitions for the diffractive event. Then,
we proceed to the computation of diffractive photoproduction of dijets
analyzing the consequences of different conventions about rapidity
gaps for this kind of process. Finally, results from the computation
are compared with available data from the ZEUS and H1 collaborations \cite{Zph,Hph}
and conclusions are presented.  \\

\noindent{\large \bf Parameterizations:}\\

It is customary to analyze diffractive DIS in terms of a so called
`diffractive structure function' $F_{2}^{D(3)}$, defined through the
triple-differential scattering cross section
\begin{equation}
\frac{d^3 \sigma^{D}}{d\beta\,dQ^2\,dx_{\pp}} \equiv \frac{4\pi
\alpha^2}{\beta\,Q^4}\left( 1-y+\frac{y^2}{2}\right) F_2^{D
(3)}(\beta,Q^2,x_{\pp}) \, ,
\end{equation} 
in the usual kinematical variables
\begin{equation}
Q^2=-q^2 \,\,\,\,\,\,\,\, x_{\pp}=\frac{q \cdot(P-P')}{q \cdot P}
\,\,\,\,\,\,\,\, \beta = \frac{Q^2}{2q \cdot (P-P')}=\frac{x}{x_{\pp}}
\, , \end{equation} where $q$, $P$, and $P'$ are the momenta of the
virtual boson, of the incident proton, and of the final state proton,
respectively, and $y=Q^2/(x\, s)$. Both the cross section and the
structure function defined in this way, include an implicit
integration over a given range of the variable $t=(P-P')^2$, and in
order to identify diffractive events, different selection criteria are
commonly used. From a theoretical point of view, the most
straightforward method is the positive identification of a final state
proton carrying a substantial fraction of the incoming proton
momentum. However, and mainly for technical reasons, the most widely
used method retains events with a pseudo-rapidity gap (absence of
hadronic activity) between the two hadronic systems in which the
hadronic final state is conventionally divided \cite{ABR}.

In the Regge inspired approach, the diffractive structure function is
assumed to be given by the product of the probability $f_{\pp
/p}(x_{\pp},t)$ to find a pomeron in the incoming proton, which only
depends on the variables $x_{\pp}$ and $t$, and a `pomeron structure function'
$F_2^{\pp}(\beta,Q^2)$, given by parton densities which behave
according to Altarelli-Parisi evolution equations and factorize as
ordinary parton distributions \cite{KUN}. Aside from the issue of the validity of
these assumptions, it is clear from the preceding paragraph that non
equivalent selection criteria for diffractive events can lead to the
extraction of non
equivalent pomeron probability densities and pomeron parton
distributions, eventually non universal \cite{Ellis}.

Alternatively, in reference \cite{ph}, it has been shown that the
diffractive structure function defined as in Eq.(1) just represents
the low $x_{\pp}$ limit of a more general semi-inclusive process,
which in perturbative QCD can be rigorous and thoroughly described
using fracture functions. 

In leading order, fracture functions account for target fragmentation 
components in the full semi-inclusive deep inelastic lepton-proton scattering cross section with an identified proton in the final state. These
components are neglected in the most usual approach based only in ordinary proton parton distributions $q^p_i(x,Q^2)$  and proton fragmentation functions $D^p_i(z,Q^2)$, which accounts for current fragmentation processes.  

Then, the full semi-inclusive cross section, 
\begin{eqnarray}
\frac{d^3 \sigma^{p/p}}{dx\,dQ^2\,dz} &=& \frac{4\pi
\alpha^2}{x\,Q^4}\left( 1-y+\frac{y^2}{2}\right)  \, \times \\
&& \left[ x\sum_i e^2_i q^p_i(x,Q^2)
\times D^p_i(z,Q^2) +  x\sum_i e^2_i M_i^{p/p}(x,z,Q^2)   \right] \, , \nonumber
\end{eqnarray} 
includes an additional contribution that can be written as a sum of hard scattering cross sections weighted by the fracture function densities corresponding to the different quark flavors $i$ for protons fragmenting into protons $M_i^{p/p}(x,z,Q^2)$, which can 
then be understood as the probabilities to find a parton of flavor $i$ in an already fragmented proton \cite{VEN}. 

In Eq.(3), the variable $z$ is, as usual, the ratio between the energy of the final state proton and that of the proton beam in the 
center of mass of the virtual photon-proton system. For very forward protons
$z\simeq 1-x_{\pp}$ and the current fragmentation components in the cross section fade away highlighting the relation between fracture functions and
diffractive or leading proton structure functions.

As for ordinary structure functions, although the scale dependence of fracture functions can be computed within perturbative QCD, they are essentially nonperturbative objects which can not be obtained from first principles in a perturbative analysis. However they can be extracted from global QCD analysis of experimental data. In reference \cite{ph} it has been shown  that a very
simple parameterization in the variables $\beta$ and $x_{\pp}$ and at an initial scale $Q^2_0=2.5 \, GeV^2$ for singlet quarks and gluons in the corresponding fracture function,
\begin{eqnarray}
xM^{p/p}_i(x_{\pp},\beta,Q^2_0) &=& N_i \,\beta^{a_i}(1-\beta)^{b_i} \, \times \\
& &\{ \beta x_{\pp}^{\alpha_i} + C_{LP} (1-\beta)^{\gamma_{LP}}[1+a_{LP}(1-x_{\pp})^{\beta_{LP}}] \} \nonumber
\end{eqnarray}
reproduces both diffractive and leading proton data as complementary
regimes of the same semi-inclusive process without need to rely in
Regge model assumptions. There, the values for the parameters were obtained fitting simultaneously diffractive and also leading proton
positron-proton DIS data from the H1 Collaboration \cite{H1D,H1LP} and can be found in reference \cite{ph}. The former data fix
the low $x_{\pp}$ behavior of the fracture function while the latter
do the same but for larger values of  $x_{\pp}$.

Regarding the diffractive selection criterion for DIS data, H1 retains events
having a rapidity gap that spans at least the region $3.3 < \eta <
7.5$, assuming that this gap definition guarantees the dominance of
single diffractive (semi-inclusive) events like $e^+(k) +p(P)
\rightarrow e^+(k')+ X + p(P')$, where $X$ is a generic hadronic final state, 
and $p(P')$ the final state proton \cite{H1D}. This assumption in fact seems to be 
justified, at least for
this kind of process and in the kinematical regime of the experiment, 
given the remarkable
agreement between the outcome of the parameterization and data
obtained by ZEUS \cite{ZD} detecting the final state proton, as pointed in
\cite{ph}. However, the assumption not necessarily holds for other gap
definitions, processes, or kinematical regimes.  In the following
section we will show that diffractive dijet photoproduction provides
an illustrative example of this situation. \\
 
\noindent{\large \bf Photoproduction}\\

Both the ZEUS and H1 Collaborations have presented results of
diffractive dijet photoproduction experiments at HERA\cite{Zphv,Zph,Hph}. These processes,
also depicted by $e^+(k) + p(P) \rightarrow e^+(k') + X + p(P')$, 
 are characterized by
the presence of at least two jets in the hadronic final state $X$ that
accompanies the outgoing proton $p(P')$ and positron $e^+(k')$.
The diffractive criterion is implemented imposing upper limits for
the pseudo-rapidity $\eta^{had}_{max}$ of the most forward particle 
belonging to the hadronic state $X$ and with energy in excess of 400 MeV  ($\eta^{had}_{max} < 3.2$  in the case of H1, while ZEUS takes $\eta^{had}_{max} < 1.8$).

The photon interaction in these processes can be either of direct or
of resolved nature, making convenient the introduction of the variable
$x_{\gamma}$, which measures the fraction of the photon momentum that
takes part in the hard interaction. The direct component of the
diffractive dijet cross section can be straightforwardly computed
replacing in the inclusive (non-diffractive) cross section, the usual
proton parton distributions by the corresponding fracture densities $xM_{p/p}^i (x_{\pp},\beta,\mu^2)$
and allowing an integration over the pertinent range in $x_{\pp}$ \cite{rph},
\begin{eqnarray}
\sigma^{direct}= \int dy\, f_{\gamma/e}(y)\int dx_{\pp} \int d\beta\,
xM_{p/p}^i (x_{\pp},\beta,\mu^2) \int d\hat{p}^2_T
\frac{d\hat{\sigma}_{i+\gamma \rightarrow k+p}}{d\hat{p}^2_T},
\end{eqnarray}
where $f_{\gamma/e}(y)$ is the flux of photons from the positron,
$\hat{p}_T$ the transverse momentum of the outgoing partons, $\mu$ the
scale at which the strong coupling constant is evaluated, and
$\hat{\sigma}_{i+\gamma \rightarrow k+p}$ the parton-photon cross
section with two partons in the final state.

For the resolved component, the procedure would be analogous to the
preceding one with the proviso that in this case no formal proof of
hard factorization guarantees it,
\begin{eqnarray}
\sigma^{resolved}= \int dy\, f_{\gamma/e}\int dx_{\gamma}
 f_{j/\gamma}\int dx_{\pp} \int d\beta \, xM_{p/p}^i \int d\hat{p}^2_T
 \frac{d\hat{\sigma}_{i+j \rightarrow
 k+p}}{d\hat{p}^2_T}. 
\end{eqnarray}
Here,
$f_{j/\gamma}(x_{\gamma},\mu^2)$ denotes parton distributions in the
photon, $\hat{\sigma}_{i+j \rightarrow k+p}$ the parton-parton cross
section with two partons in the final state. For simplicity we
have dropped the arguments for the densities and also the sums over the
partonic indexes $i$ and $j$.

In any case, before dealing with the issue of hard factorization and
making comparisons between the parameterization and the available data
sets, it is imperative to analyze the compatibility between the gap
criteria implicit in them.

Given that the profile of the hadron activity as a function of
the pseudo-rapidity of the most forward hadron, 
and thus the consequences of a given gap convention,
depends on various non perturbative hadronization effects, in what
follows we simulate them using a variant of the Monte Carlo generator
POMPYT 2.6 \cite{POM}, modified in order to include parameterizations for fracture
functions. 

The inclusion of fracture functions in POMPYT imply to abandon explicitly the pomeron flux factorization hypothesis, modifying both the shape and scale dependence of the diffractive parton densities as a function of $x_{\pp}$, as required by DIS data \cite{ph}. In what follows, all the simulations are computed using the best fit to diffractive DIS data, denoted as SET A in reference \cite{ph} as the fracture function,
and SaS 2M from reference \cite{SAS} for the photon.

\setlength{\unitlength}{1.mm}
\begin{figure}[hbt]
\begin{picture}(100,80)(0,0)
\put(13,-70){\mbox{\epsfxsize12.0cm\epsffile{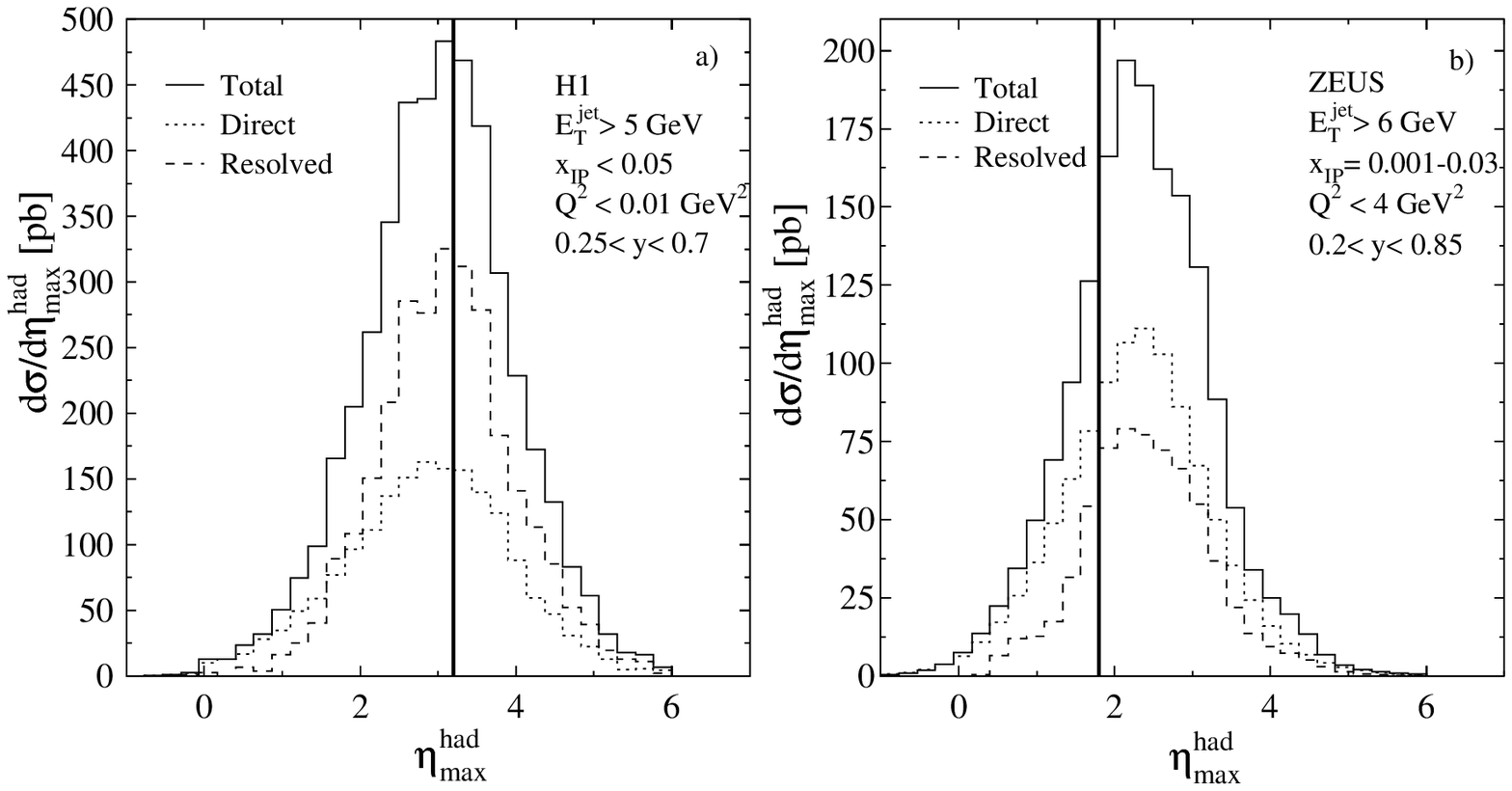}}}
\put(10,9){\mbox{\footnotesize \bf Figure 1:} {\footnotesize Dijet photoproduction events generated in the kinematical regimes}} 
\put(10,5){\mbox{\footnotesize of the {\bf a)} H1  and {\bf b)} ZEUS experiments}}
\end{picture}
\end{figure}

In Figure (1), dijet photoproduction events generated in the
kinematical regimes of  H1 (Fig. 1a) and ZEUS (Fig. 1b) are shown as a function of $\eta^{had}_{max}$. The thick vertical lines represent the upper cuts in
$\eta^{had}_{max}$ applied by each collaboration.

It is worthwhile noticing that even for the same kind of process, the
different kinematical regimes covered by both experiments given for
example by cuts in the transverse energy $E_T$ and pseudo-rapidity
$\eta$ of the two most energetic jets, etc., yield rather different
distributions in $\eta^{had}_{max}$ and thus, are affected in a
different way even for the same pseudo-rapidity gap definition. For
ZEUS, approximately 25 \% of the total number of events are
concentrated at $\eta^{had}_{max}< 1.8$, while for H1 kinematics
approximately 10 \% of them survives the same constraint.

In the computation, an additional multiplicative normalization factor
has been applied to the fracture function in order to take into
account the fact that the original parameterization was adjusted to
data already filtered by a gap definition. This
normalization factor can be defined in a first approximation as the ratio
between the total number of dijet photoproduction events and those
that satisfy the gap definition implicit in diffractive DIS data, both
quantities computed in the kinematical range where the simulation 
takes place. In the present case, this rate is found to be 1.14 for
ZEUS and 1.61 for H1.

\setlength{\unitlength}{1.mm}
\begin{figure}[hbt]
\begin{picture}(100,130)(0,0)
\put(13,-15){\mbox{\epsfxsize12.0cm\epsffile{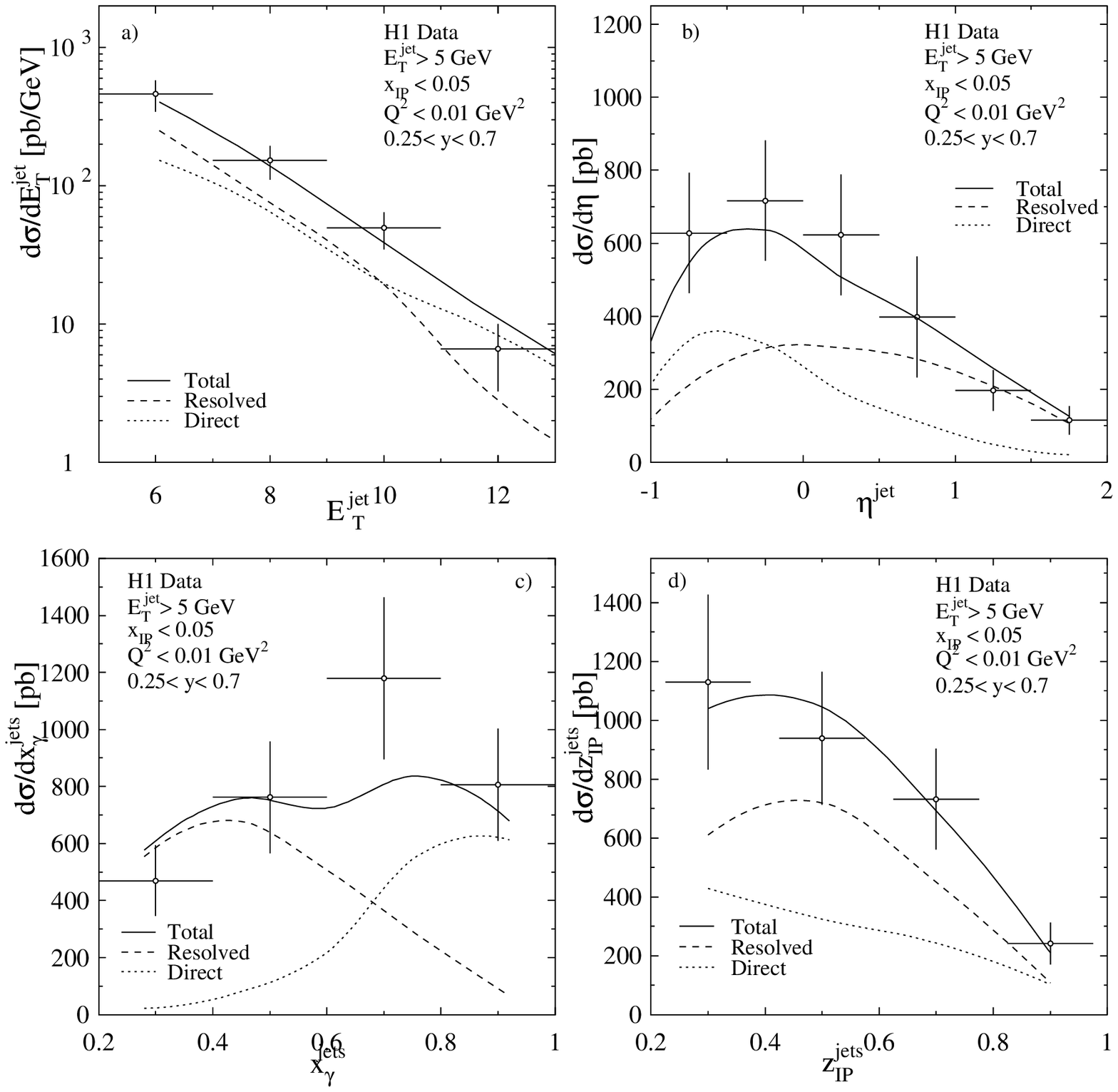}}}
\put(5,1){\mbox{\footnotesize \bf Figure 2:} {\footnotesize  Comparison between H1 dijet photoproduction data and the outcome}} 
\put(5,-3){\mbox{\footnotesize   the fracture function parameterization.}}
\end{picture}
\end{figure}

Figures (2) and (3) show the comparison between dijet photoproduction
data \cite{Zph,Hph} and the outcome of the fracture function parameterization obtained 
from DIS. The cross sections are computed in picobarns and as functions of the hadronic level variables defined within each experiment.

As it can be seen, the agreement is quite good for both sets of data 
although the effects of taking into account the different rapidity gap conventions are large. It is important to notice
that neglecting the stringent restriction applied in the ZEUS
measurement ($\eta^{had}_{max}< 1.8$), the computation would have led
to cross sections up to five times larger. 

\setlength{\unitlength}{1.mm}
\begin{figure}[hbt]
\begin{picture}(100,130)(0,0)
\put(13,-15){\mbox{\epsfxsize12.0cm\epsffile{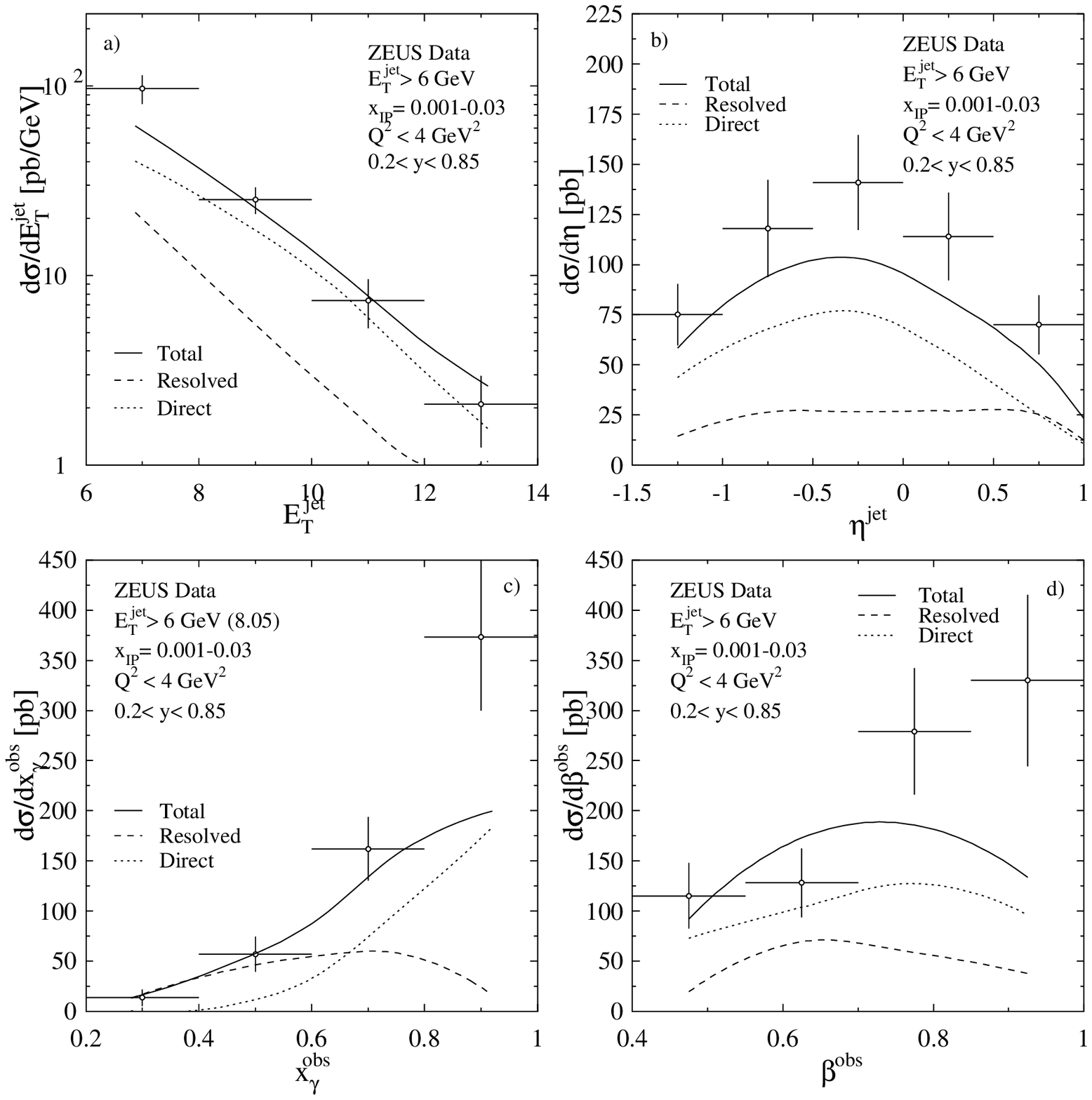}}}
\put(5,1){\mbox{\footnotesize \bf Figure 3:} {\footnotesize  Comparison between ZEUS dijet photoproduction data and the outcome}} 
\put(5,-3){\mbox{\footnotesize   the fracture function parameterization.}}
\end{picture}
\end{figure}

The comparison with ZEUS data shows that the numerical results tend to underestimate the cross sections for large values of $\beta^{obs}$. This fact
could indicate the need of a further adjustment in the diffractive parameterization
in that region, which in fact is allowed by the present level of uncertainties  in DIS data, specially at lower scales.   

For H1 photoproduction data, the 
gap definition is very similar to that of diffractive DIS data so the
effect of neglecting at all such cuts is relatively small when analyzing event distributions in variables other than $\eta^{had}_{max}$.
However, if one wants to reproduce the whole picture including the  $\eta^{had}_{max}$-dependence, of course they can not be neglected and some care must be taken in order to not apply the cuts twice, as pointed by the large normalization factor.

In the  precedent discussion, specifically in the
computation of the normalization factor, we have assumed that the
rapidity gap restriction only reduces uniformly the number of events,
i.e., without changing the distribution of events in other kinematical
variables. This seems to be a good approximation for the $E_T$ and $\eta$ distributions,
but it is not for $x_{\pp}$. The Monte Carlo simulation shows a clear correlation
between the cuts applied in $\eta^{had}_{max}$ and the mean for the
distributions in $x_{\pp}$ implying that the larger is the gap, the
lower is the mean $x_{\pp}$ proved, as shown in figure (4).

\setlength{\unitlength}{1.mm}
\begin{figure}[hbt]
\begin{picture}(100,78)(0,0)
\put(13,-76){\mbox{\epsfxsize12.0cm\epsffile{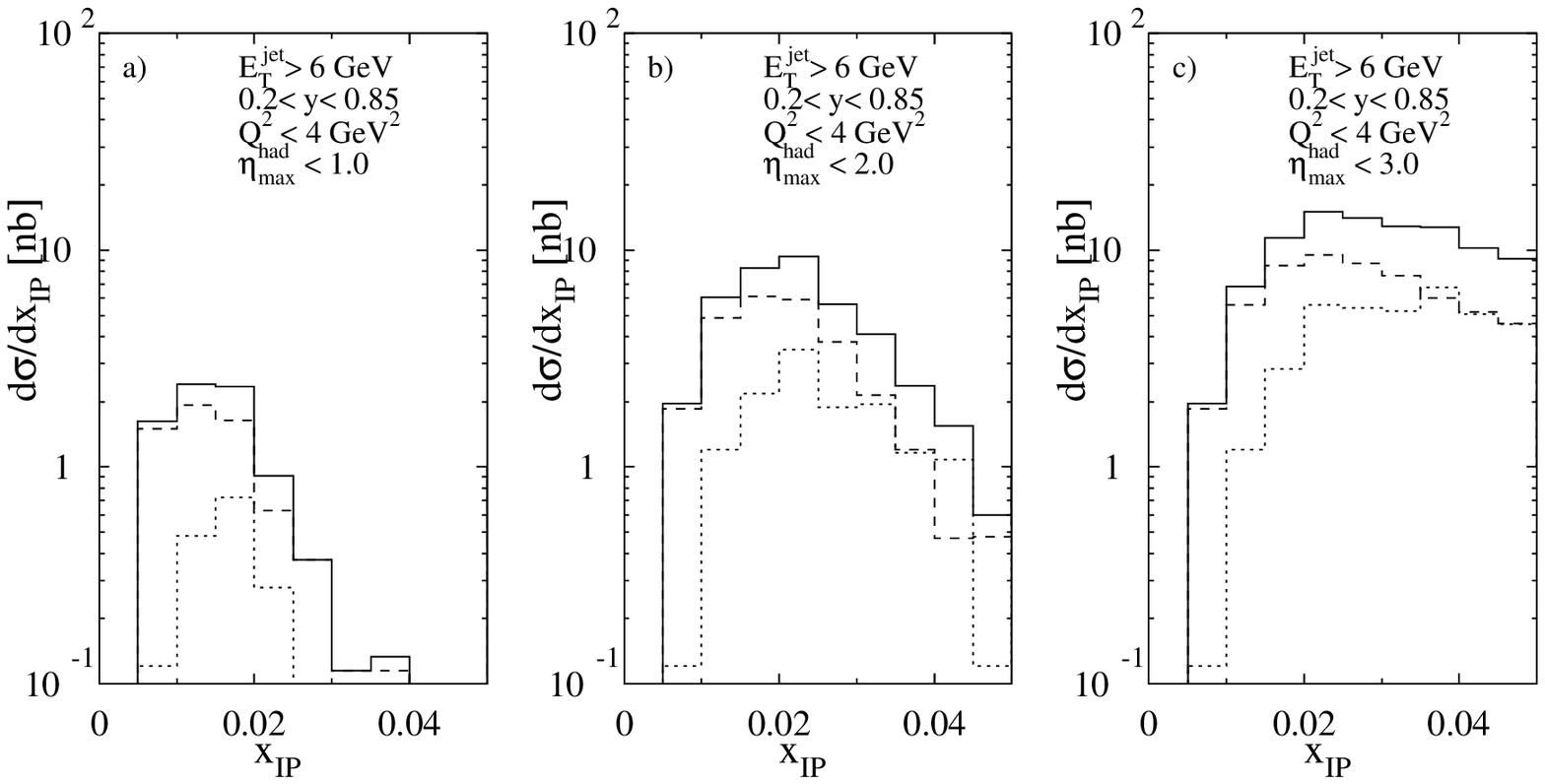}}}
\put(8,5){\mbox{\footnotesize \bf Figure 4:} {\footnotesize Distribution of dijet events as a function $x_{\pp}$ for different rapidity }} 
\put(8,1){\mbox{\footnotesize gap definitions ({\bf a)} $\eta^{had}_{max}< 1.0$, {\bf b)} $\eta^{had}_{max}< 2.0$, and  {\bf c)} $\eta^{had}_{max}< 3.0$) }}
\end{picture}
\end{figure}

The direct consequence of the above mentioned effect is that different
gap definitions weight differently contributions coming from a given range of
$x_{\pp}$, thus adding to the discrepancy between diffractive structure
functions extracted from different experiments. In our approach, this
inconvenience has been straightforwardly overcame computing the  normalization factor for various bins in $x_{\pp}$. The main effect of this procedure in the comparison with H1 data is to enhance the cross section at low $z_{\pp}$ (large $x_{\pp}$) in figure (3d) and reduce it slightly at
large $z_{\pp}$. For ZEUS, the normalization effect is small so the correction
is almost negligible. 

This considerations about rapidity gap definitions apply also to the comparison between diffractive DIS and 
diffractive dijet production within proton-antiproton collisions at Tevatron, 
where large discrepancies have been observed \cite{COLLB,GOUE}. In the comparison of DIS data 
and these processes, the effects associated with the gap definition can even 
be enhanced not only because of the differences in the kinematical ranges covered  but also in the extrapolation of the diffractive DIS parameterizations to rather large values $x_{\pp}$, dominant in proton-antiproton collisions.\\

\vspace*{3mm}
\noindent{\large \bf Hard Factorization}\\

Regarding the issue of hard factorization, it is worth noticing that in both ZEUS and H1 photoproduction experiments, and even in the regions dominated by the contributions coming from resolved photons ($x_{\gamma} < 0.7$), no hints of significant factorization breaking effects are found, confirming our
initial assumption about hard factorization in these kind of
processes. This is particularly apparent for H1 data which contains a much larger fraction of events initiated by resolved photons and would indicate
that hard factorization, if not exact, is a very good approximation. 

It is also interesting to notice that at variance with the analysis of H1 data of reference \cite{Hph}, in the present treatment no substantial overestimate of the cross section as a function $z_{\pp}$ is found. In reference \cite{Hph} this
excess was interpreted of a reduced gap survival probability \cite{BJ}, and an ad hoc correction factor was fixed with the data. 
In the present approach these discrepancies can be traced back to the non equivalence of rapidity gap definitions, which encompasses effects related to 
spectator interactions, but in a more predictable way. \\

\noindent{\large \bf Conclusions:}\\

Dijet diffractive photoproduction has been analyzed in the framework of 
fracture functions finding that the main effects that conspire against 
a unified picture for diffractive DIS and diffractive photoproduction
in terms of diffractive parton distributions are related to non-equivalent 
criteria used to select diffractive events.

The connection between different rapidity gap definitions is 
model-dependent in the sense that it includes non-perturbative 
hadronization details. However, using a slight modification of the POMPYT 
Monte Carlo generator, devised in order to include in it fracture functions,
a consistent and accurate picture can be drawn. 

In this way, the analysis suggests that the supposed discrepancies between the 
diffractive parameterizations coming from these experiments are rather due to 
the non-universal character of rapidity gaps, than to a hard 
factorization breakdown, and that the potential factorization breaking mechanisms are negligible, well beyond the accuracy of the present data.\\

\noindent{\large \bf Acknowledgements}

I would like to thank D. de Florian and C. A. Garc\'{\i}a Canal for interesting comments and suggestions.\\

\pagebreak

\end{document}